# Nanoscale Conductivity Imaging of Correlated Electronic States in WSe$_2$/WS$_2$ Moiré Superlattices

Zhaodong Chu[1†], Emma C Regan[2,3,4†], Xuejian Ma[1], Danqing Wang[2,3,4], Zifan Xu[1], M. Iqbal Bakti Utama[2,4,5], Kentaro Yumigeta[6], Mark Blei[6], Kenji Watanabe[7], Takashi Taniguchi[8], Sefaattin Tongay[6], Feng Wang[2,4,9*], Keji Lai[1*]

[1]Department of Physics, The University of Texas at Austin, Austin, Texas 78712, USA

[2]Department of Physics, University of California at Berkeley, Berkeley, California 94720, USA

[3]Graduate Group in Applied Science and Technology, University of California at Berkeley, Berkeley, California 94720, USA

[4]Material Science Division, Lawrence Berkeley National Laboratory, Berkeley, California 94720, USA

[5]Department of Materials Science and Engineering, University of California at Berkeley, Berkeley, California 94720, USA

[6]School for Engineering of Matter, Transport and Energy, Arizona State University, Tempe, Arizona 85287, USA

[7]Research Center for Functional Materials, National Institute for Materials Science, 1-1 Namiki, Tsukuba 305-0044, Japan

[8]International Center for Materials Nanoarchitectonics, National Institute for Materials Science, 1-1 Namiki, Tsukuba 305-0044, Japan

[9]Kavli Energy NanoSciences Institute at University of California Berkeley and Lawrence Berkeley National Laboratory, Berkeley, California 94720, USA

† These authors contributed equally to this work

*Corresponding emails: fengwang76@berkeley.edu, kejilai@physics.utexas.edu

# Abstract

We report the nanoscale conductivity imaging of correlated electronic states in angle-aligned $WSe_2/WS_2$ heterostructures using microwave impedance microscopy. The noncontact microwave probe allows us to observe the Mott insulating state with one hole per moiré unit cell that persists for temperatures up to 150 K, consistent with other characterization techniques. In addition, we identify for the first time a Mott insulating state at one electron per moiré unit cell. Appreciable inhomogeneity of the correlated states is directly visualized in the hetero-bilayer region, indicative of local disorders in the moiré superlattice potential or electrostatic doping. Our work provides important insights on 2D moiré systems down to the microscopic level.

The vertical stacking of two-dimensional (2D) materials with similar lattice constants at small twist angles produces long-wavelength moiré superlattices, which can substantially modify the electronic structures of individual layers[1-3]. As exemplified by the groundbreaking experiments[4-6] and subsequent works[7-12], a plethora of exotic behaviors such as Mott insulating states[4-9], superconductivity[5,7-9], magnetic phases[10-12], and quantum anomalous Hall effect[12] have been reported on twisted bilayer graphene near the magic angle and trilayer graphene/boron nitride moiré superlattices. Moiré systems composed of transition metal dichalcogenide (TMD) heterostructures offer a complementary and equally important material platform to investigate the rich many-body phenomena[13]. Because of the strong spin-orbit coupling, only two-fold degeneracy remains in the TMD-based moiré systems, giving rise to an ideal playground to simulate the Mott-Hubbard physics in 2D triangular lattices[13-18]. To date, evidence of charge ordering and correlated states at partial moiré band fillings has been reported in both homo-bilayer[16] and hetero-bilayer TMD superlattices[17,18], with many other novel electronic states expected to emerge[19].

A notable experimental challenge in the investigation of TMD moiré superlattices lies in the large metal-semiconductor contact resistances, especially at the low-density regime where correlation effects manifest themselves. As a result, optical measurements such as reflectance[17] and optically detected resistance and capacitance (ODRC)[18] are preferred over traditional transport techniques. A non-contact method that directly measures the electrical conductivity is therefore desirable to understand the insulating nature of Mott-like states in these materials. Moreover, as electrical properties of moiré superlattices are highly sensitive to the local atomic arrangement and disordered effects in the 2D sheets[20-24], it is imperative to study the correlation physics in a spatially resolved manner. In this work, we report the optical study and nanoscale conductivity imaging of angle-aligned $WSe_2/WS_2$ moiré hetero-bilayers by combining ODRC and microwave impedance microscopy (MIM) on the same devices. At sufficiently low temperatures, while the overall gate dependence of local conductivity in the monolayer regions are described by the ambipolar field effect, pronounced resistivity peaks on both electron and hole sides are observed in the superlattice area at certain moiré band fillings. The resistivity peak at one hole per lattice site is consistent with the Mott insulator state reported previously[17, 18]. Resistivity peak at one electron per lattice site, on the other hand, is reported for the first time, which indicates that the Mott state is also present in electron-doped $WS_2/WSe_2$ moiré systems. In addition, the microwave images show clear spatial inhomogeneity of the correlated states, presumably due to local disorders in the moiré potential or electrostatic doping. Our results provide important insights for improving the quality of TMD moiré superlattices towards new discoveries in these fascinating material systems.

The samples in our study are WSe$_2$/WS$_2$ hetero-bilayers, which are encapsulated by top and bottom hexagonal boron nitride (hBN) dielectrics, as illustrated in Fig. 1a. Few-layer graphene (FLG) layers are used as the gate and contact electrodes (details in Supplementary Information S1). The twist angle between the two monolayers is $\theta \sim 60°$ (Supplementary Information S2), resulting in a moiré superlattice constant $L_M = a/\sqrt{\delta^2 + (\pi/3 - \theta)^2} \sim 8$ nm, where $\delta \sim 4\%$ is the lattice mismatch between monolayer WSe$_2$ ($a = 0.328$ nm) and WS$_2$ flakes. As seen in the optical image in Fig. 1b, half of the heterostructure (Region I) is covered by the top graphene layer, whereas the other half (Region II) is not. Significant inhomogeneity due to trapped bubbles are observed in this heterostructure devices. The optical absorption spectrum on Region II (Fig. 1c) shows that the WSe$_2$ A exciton is split into three peaks, indicative of the presence of the moiré potential[25] (more data in Supplementary Information S3).

To study the correlated electronic behavior in the sample, we perform ODRC measurements at 10 K. A d.c. voltage applied to the top gate ($V_{tg}$) is used to tune the carrier concentration in region I, and an a.c. voltage ($\Delta \tilde{V}$) induces charge redistribution between regions I and II. A probe laser in resonance with the lowest-energy WSe$_2$ exciton peak is used to measure charge flow in the sample[18] (details in Supplementary Information S4). Fig. 1d shows the measured ODRC signals as a function of $V_{tg}$ at three excitation frequencies. Consistent with the previous report[17,18], a prominent dip in the change of optical contrast is observed at $V_{tg} \sim -1.37$ V, corresponding to a filling factor $n/n_0 = -1$ or one hole per moiré unit cell ($n$ represents the charge concentration and $n_0 = 1/[L_M^2 \sin(\pi/3)] \sim 1.8 \times 10^{12}$ cm$^{-2}$ is the density of one charge per moiré unit cell). In addition to this Mott state, a satellite dip also appears at factional filling $n/n_0 = -1/3$.

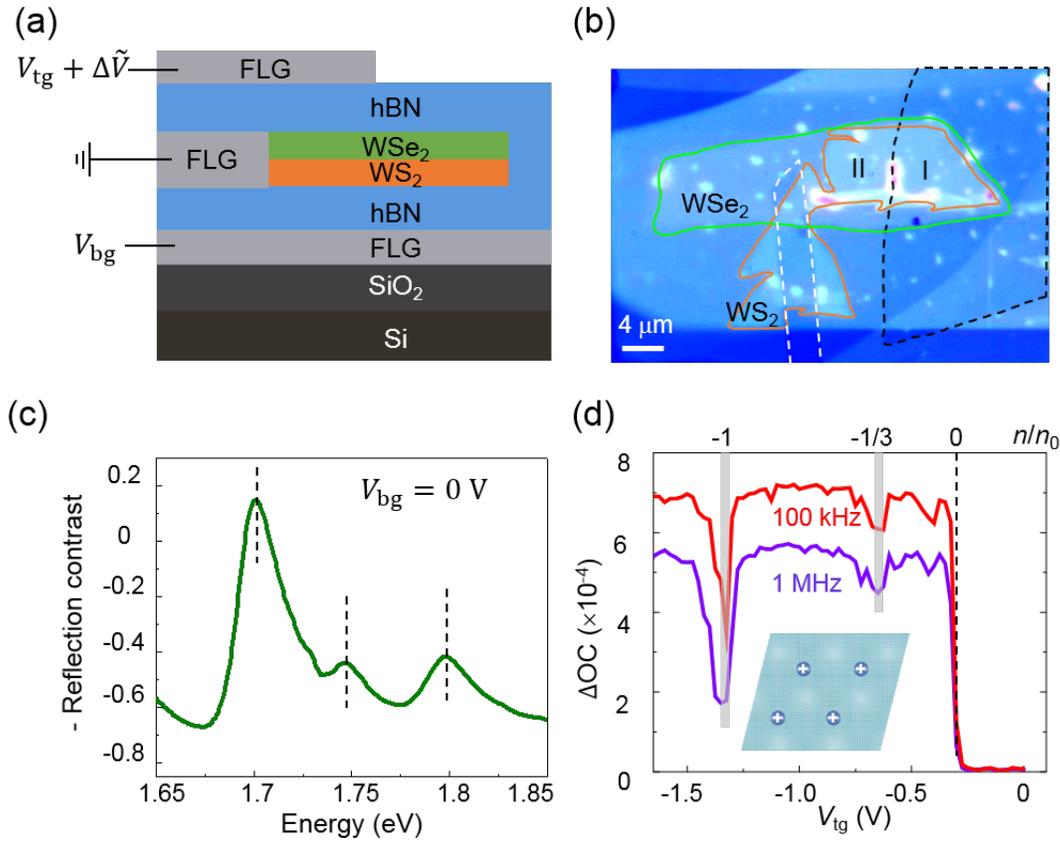

FIG. 1. (**a**) Schematic of the gated $WSe_2/WS_2$ superlattice device structure. The thicknesses of top and bottom hBN are 11 nm and 16 nm, respectively. (**b**) Optical microscopy image of the heterostructure device. The $WSe_2$, $WS_2$ monolayer flakes, FLG contact and top gate are outlined in green, orange, white, and black, respectively. (**c**) Optical absorption spectrum in Region II of the heterostructure in (b). The three peaks indicated by dashed lines are due to splitting of the $WSe_2$ A exciton, which is indicative of intra-layer moiré excitons in the $WSe_2/WS_2$ superlattice. (**d**) ODRC signals at two different a.c. excitation frequencies as a function of the top gate voltage and filling factors. The dashed line denotes the charge-neutral point (~ -0.3V) and the shaded areas correspond to the Mott insulator state at $n/n_0 = -1$ and correlated insulating state at -1/3. The inset illustrates the Mott insulator state with one hole per moiré unit cell.

The nanoscale conductivity of the moiré superlattice device is studied by microwave impedance microscopy (MIM)[26], a versatile technique for imaging the spatial uniformity of electrical properties in advanced materials[27-30]. As shown in Fig. 2a, a microwave signal (~ 10 µW at 1 GHz) is delivered to a chemically etched tungsten tip, which is mounted on a quartz tuning fork (TF) for distance feedback[31,32]. The two MIM output signals, proportional to the imaginary (MIM-Im) and real (MIM-Re) parts of the tip-sample admittance, are demodulated at the TF resonant frequency (~ 37 kHz). The spatial resolution is on the order of 100 nm, which is comparable to the tip diameter, as shown in the scanning electron microscopy (SEM) image in Fig.

2a. The MIM response to the sheet conductance of a buried 2D layer can be modeled by finite element analysis (FEA)[26]. Fig. 2b shows the simulated MIM result as a function of the 2D conductance σ of, e.g., the WS$_2$ monolayer in both semi-log (main plot) and linear (inset) scales. Note that the MIM-Im signal increases monotonically with increasing σ, whereas MIM-Re peaks at σ ~ 10$^{-7}$ S·sq (see details in Supplementary Information S5). For comparison, the measured room-temperature MIM signals as a function of the back-gate voltage ($V_{bg}$) in the monolayer WS$_2$ region are shown in Fig. 2c. Below a threshold voltage of ~ 0.3 V, MIM signals in both channels are negligible, indicating the absence of mobile carriers when the Fermi level is inside the semiconducting gap. For $V_{bg}$ > 0.3V, the electrostatic doping occurs, as evidenced by the increase of MIM signals. The close resemblance between the simulated (Fig. 2b inset) and measured (Fig. 2c) results suggests that the device behaves as a regular back-gated field-effect transistor (FET). For simplicity, we will in the following only present the MIM-Im data because of its monotonic relationship with respect to the local conductance. A complete set of the MIM data are found in Supplementary Information S6. Similar results on a different WSe$_2$/WS$_2$ sample with near-0° twist angle are also included in Supplementary Information S7.

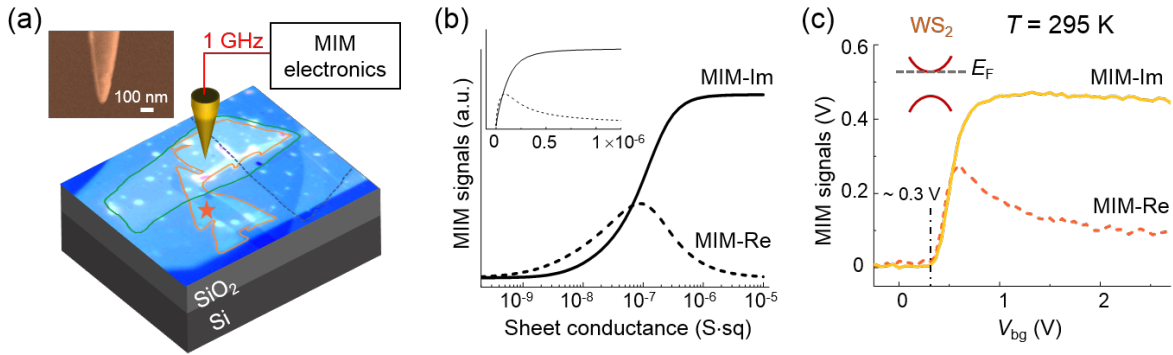

FIG. 2. (**a**) Schematic diagram of the TF-MIM measurements on the WSe$_2$/WS$_2$ hetero-bilayer device. The inset shows the SEM image of the etched tungsten tip. (**b**) FEA simulated MIM signals as a function of the sheet conductance of the buried TMD layers. The inset shows the same plot in the linear scale. (**c**) Measured room-temperature MIM signals on the WS$_2$ monolayer (orange star in **a**) as a function of the back-gate voltage. The black dash-dotted line indicates the charge-neutral point for electrons at ~ 0.3 V, where the Fermi level resides at the conduction band minimum (inset).

The cryogenic MIM experiment is carried out in a helium-flow cryostat with a base temperature ~ 20 K. Fig. 3a displays the sample topography taken by the TF feedback, where some surface impurities or air bubbles are clearly visible. The simultaneously acquired MIM-Im image at $V_{bg}$ = 0 V and T = 24 K (Fig. 3b) shows that only the contact and top-gate FLG regions are conductive. Strikingly, while the individual TMD layers and the heterostructure (HS) region are both highly insulating at zero $V_{bg}$, they exhibit very different gate dependences. As seen in the inset of Fig. 3b, monolayer WSe$_2$ is conductive at $V_{bg}$ = -2 V due to the electrostatic doping of holes. The HS region, however, is strongly insulating. Such an unusual behavior of the WSe$_2$/WS$_2$ bilayer is beyond the simple description of a back-gated FET and worth further investigation.

Since the vertically stacked layers are relatively fragile and could be easily damaged by repeated scans, we focus on point measurements for systematic studies of the gate dependence. Fig. 3c shows the MIM-Im response on three points (WS$_2$, WSe$_2$, and HS) indicated in Fig. 3a when $V_{bg}$ is swept from -2.7 to 2.7 V. The ambipolar-like FET behavior is observed in both monolayers. We notice that the WS$_2$ layer is much more difficult to turn on with hole doping, similar to previous FET studies[33-36]. In the HS region, on the other hand, two prominent drops of MIM-Im signals appear at around $V_{bg}$ ~ -2.1 V and +1.8 V. On the hole-doping side, given the threshold voltage ~ -0.7 V for injecting holes into the WSe$_2$ layer, the insulating state corresponds to a filling factor $n/n_0 = -1$, i.e. one hole per moiré unit cell. The results are also consistent with the ODRC data (Fig. 1d) and previous investigations[17,18]. The difference in the voltage values is because the ODRC measurements use the top gate with the top hBN thickness of 11 nm and the MIM measurements use the bottom gate with the bottom hBN thickness of 16 nm.

The insulating state with electron doping, however, has not yet been observed before. Given the ~ 0.3 V threshold voltage for injecting electrons, the observed insulating state at $V_{bg}$ ~ 1.8 V corresponds to filling factor $n/n_0 = +1$, i.e. one electron per moiré unit cell. The fact that both WSe$_2$ and WS$_2$ are turned on at this threshold voltage suggests that the gate-induced electrons are located in both layers. Interestingly, while the two monolayer regions display comparable MIM-Im signals in the ambipolar FET curves, the overall MIM signal level on the electron-doped side is considerably lower than that on the hole-doped side, indicative of a large intrinsic resistance of the former. Such a property may lead to large contact resistance, which represents a big challenge for conventional transport measurements to access the electron side of WSe$_2$/WS$_2$ moiré superlattices at low doping and low temperatures. Further experiments are needed to examine this hypothesis. We emphasize that the requirement of good Ohmic contacts is much less stringent for the MIM measurements, which enables the observation of Mott states on both electron and hole sides. The Mott states persist up to ~ 150 K and eventually vanish at 295 K. A rough estimate of the thermal activation gap of the Mott state is thus 10 ~ 20 meV, consistent with the earlier reports[17,18].

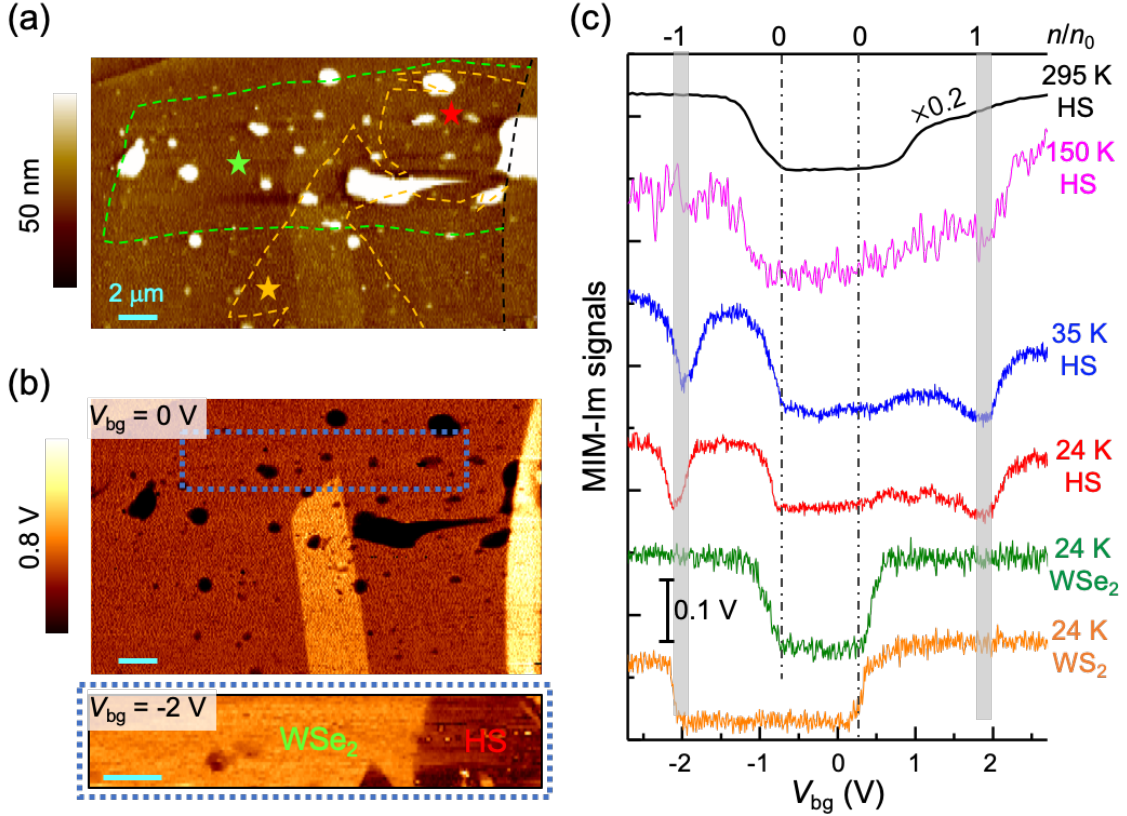

FIG. 3. (**a**) Topographic image of the WSe$_2$/WS$_2$ device taken by the tuning-fork feedback. The WSe$_2$, WS$_2$, and top FLG regions are labeled by green, orange, and black dashed lines, respectively. The green, orange, and red stars mark the locations where the MIM curves in (d) are taken. (**b**) MIM-Im image at $T = 24$ K and $V_{bg} = 0$ V in the same area as (a). Inset: MIM-Im images in the blue dashed box in (b) taken at $V_{bg} = -2$ V. Scale bars in (a – b) are 2 μm. (**c**) MIM-Im signals at various locations and temperature. The two dash-dotted lines denote the charge neutral points, -0.7 V for holes in WSe$_2$ and +0.3 V for electrons in WS$_2$. The shaded areas correspond to the Mott insulating states at $n/n_0 = \pm 1$. The data at 295K HS is scaled by 0.2 for compact visualization.

Finally, we briefly discuss the spatial variation of the Mott insulating state in the moiré superlattice. Figs. 4a and 4b show the close-up AFM and MIM-Im images near the HS region at $T = 24$ K and $V_{bg} = -2$ V. In addition to the surface impurities visible from the topography, inhomogeneous local conductivity is clearly observed in the MIM image. In Fig. 4c, the MIM-Im signals when $V_{bg}$ is swept from -2.7 V to 2.7 V are plotted for the three points labeled in Fig. 4b. The dips for $n/n_0 = \pm 1$, although present in all curves, differ significantly in their strengths and positions with respect to $V_{bg}$. The electrical inhomogeneity is indicative of spatial disorder in the correlated insulating state, which may be caused by strains, defects, density inhomogeneity from electrostatic gating, and local variation of the twist angle. Future improvement on the sample uniformity is thus critical for in-depth understanding of correlation physics in 2D moiré superlattices.

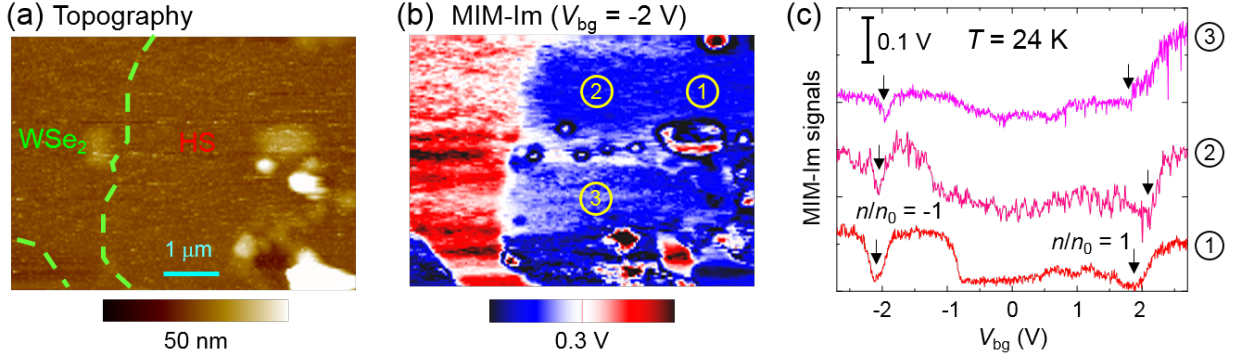

FIG. 4. (**a**) Topography and (**b**) MIM-Im images at $V_{bg}$ = -2 V for a close-up view in the heterostructure region. (**c**) MIM-Im signals at three different locations labeled in (a) as a function of the back-gate voltage. The positions of $n/n_0 = \pm 1$ are indicated by arrows for each curve.

In summary, we report the near-field microwave imaging of correlated states in semiconducting TMD hetero-bilayer moiré superlattices. The ability to probe local conductivity in buried structures allows us to observe the Mott insulating phases with both hole and electron doping at temperatures up to 150 K. The nanoscale microwave images reveal substantial spatial inhomogeneity in the correlated states. Our work provides a guideline for improving the material quality towards novel electronic states in 2D moiré systems.

The work at UT-Austin (Z.C., X.M. and Z.X., and K.L.) was supported by the U.S. Department of Energy (DOE), Office of Science, Basic Energy Sciences, under Award No. DE-SC0019025. The MIM instrumentation is supported by the Welch Foundation Grant F-1814. The ODRC measurements are supported by the Director, Office of Science, Office of Basic Energy Sciences, Materials Sciences and Engineering Division of the US Department of Energy under contract number DE-AC02-05CH11231 (van der Waals heterostructures program, KCWF16). The heterostructure fabrication is supported by the US Army Research Office under MURI award W911NF-17-1-0312. K.W. and T.T. acknowledge support from the Elemental Strategy Initiative conducted by the MEXT, Japan, Grant Number JPMXP0112101001, JSPS KAKENHI Grant Number JP20H00354 and the CREST (JPMJCR15F3), JST. S.T acknowledges support from DOE-SC0020653, NSF DMR 1552220, DMR 1904716, and NSF CMMI 1933214. E.C.R.

acknowledges support from the Department of Defense (DoD) through the National Defense Science & Engineering Graduate Fellowship (NDSEG) Program.

## S1. Fabrication details of the WSe$_2$/WS$_2$ moiré superlattices device.

The WSe$_2$/WS$_2$ heterostructure devices were fabricated by using a dry-transfer method with a polyethylene terephthalate (PET) stamp[S1]. First, we exfoliated and transferred the TMD monolayers, few-layer graphene and thin hBN onto Si substrates with 90 nm SiO$_2$. Secondly, a PET stamp was used to sequentially pick up the few-layer graphene top gate, top hBN flake, the WSe$_2$ monolayer, the WS$_2$ monolayer, the few-layer graphene contact, the bottom hBN flake and the few-layer graphene back gate. Before picking up WS$_2$, the PET stamp was reoriented to ensure a near-60° angle between the WSe$_2$ and WS$_2$ flakes. All the layers were then stamped onto a clean SiO$_2$/Si substrate. The PET and samples were heated to 60 °C during the pick-up and to 130 °C for the stamping process. Finally, the PET was dissolved in dichloromethane overnight at room temperature. The gold contacts (~75 nm with a ~5-nm-thick chromium adhesion layer) to the few-layer graphene flakes were made using electron-beam lithography and electron-beam evaporation.

## S2. Determination of the twist angle between monolayer WSe$_2$ and WS$_2$ flakes.

We determined the twist angle (60.0 ± 0.3 degrees) between WSe$_2$ and WS$_2$ flakes by performing polarization dependent second harmonic generation (SHG) measurements on both the TMD monolayers and heterostructure regions in the device[S2], [S3] (Fig. S2). The absence of SHG signal on the heterostructure region indicates that the twist angle is close to 60 degrees, rather than 0 degree.

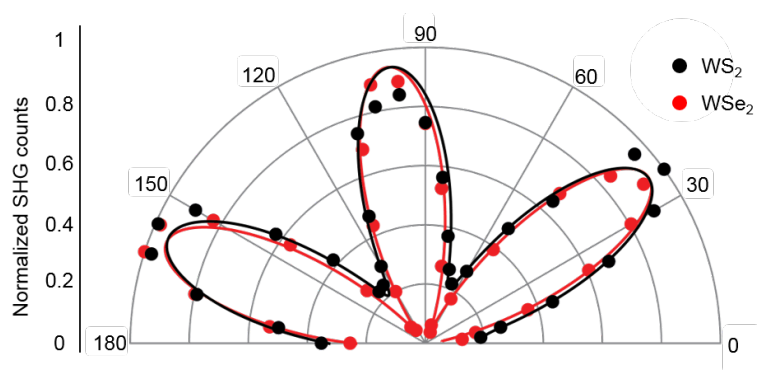

Fig. S2. Polarization-dependent SHG signal on monolayer WSe$_2$ (red circles) and WS$_2$ (black circles) of the device and corresponding fittings (red and black curves). The twist angle between the WSe$_2$ and WS$_2$ flakes is 60.0 ± 0.3 degrees.

**S3. Back gate dependent absorption spectra.**

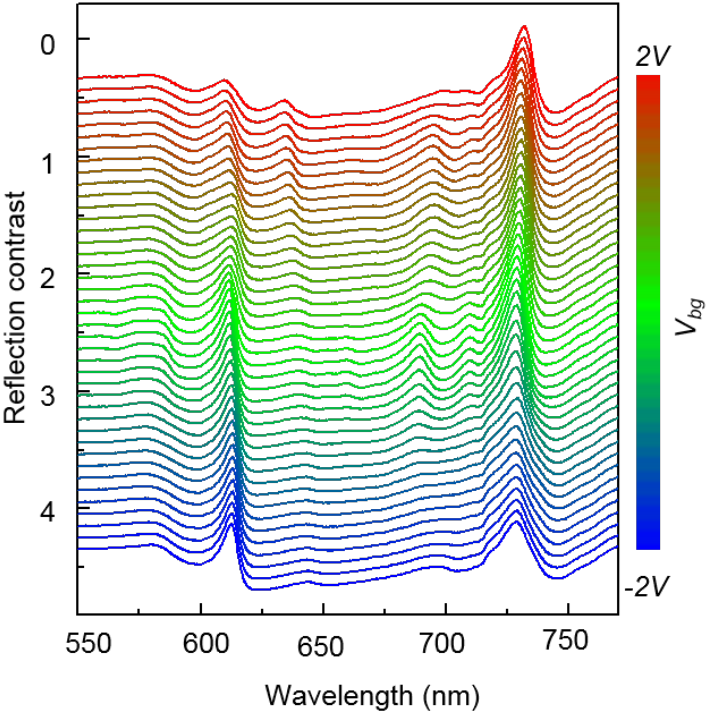

Fig. S3. Absorption spectra of the back-gate-only region (Region II in Fig. 1b in the main text) at 10 K while tuning the back-gate voltage from -2 V to 2 V with 0.1 V increment.

## S4. ODRC measurements at 10 K.

The ODRC measurements on the WSe$_2$/WS$_2$ moiré superlattice were performed following the experimental description in reference [S4]. The superlattice devices include a local top gate: region I is covered by the gate, and region II is not. The few-layer-graphene top gate carries both d.c. voltage ($V_{tg}$) and a.c. excitation voltage ($\Delta \tilde{V}$). The d.c. voltage is used to continuously control the carrier doping in region I, while a small (10 mV) a.c. voltage with high excitation frequency ($w$) is applied to redistribute the charge between region I and region II without any changes on the total charge. The charge redistribution depends on the quantum capacitance ($C_Q$) and resistance ($R$) of the moiré superlattice in region I, and leads to a change of carrier concentration ($\Delta \tilde{n}$) in region II, which can be detected optically. We directly measured the change in optical contrast at region II, $\Delta OC$, using a laser probe (1.70 eV) resonant with the lowest energy WSe$_2$ A exciton peak. As shown in Fig. S4a, the relationship between $\Delta \tilde{V}$, $\Delta \tilde{n}$ and $\Delta OC$ can be described by an effective circuit model (Eq. S1) [S4].

$$\Delta OC = \alpha \Delta \tilde{n} = \frac{\alpha}{A_2 e} \Delta \tilde{V} \frac{C_{1t}}{C_{1t}+C_{1b}} \frac{1}{\frac{1}{C_{1t}+C_{1b}}+\frac{1}{C_{2b}}+\frac{1}{C_Q}+iwR} \quad (S1)$$

Here, α is the optical detection responsivity in region II, $C_{1t}$ and $C_{1b}$ are the geometric capacitances between the TMD and the top and bottom gates in region I respectively. $C_{2b}$ is the TMD-bottom gate capacitance in region II. These geometric capacitances $C_i$ ($i = 1t, 1b, 2b$) are set by $C_i = \varepsilon_0 \varepsilon_r A_i / d_i$, where $\varepsilon_r$ is the dielectric constant ($\sim 4.2 \pm 0.4$) of the gate dielectric, i.e. hBN, $A_i$ and $d_i$ are the relevant capacitor area and separation, respectively.

Besides the result shown in Figure 1d in the main text, Fig. S4b-c show two additional rounds of measured ODRC signals as a function of the top-gate voltage ($V_{tg}$) in region I at various excitation frequencies. The locations of the probe laser spot might have slightly shifted between the three rounds of measurements. When $V_{tg} > -0.3$ V, the heterostructure is charge neutral and the ODRC signal is negligibly small. When $V_{tg} < -0.3$ V, hole doping starts in region I and charge redistribution occurs, leading to a large increase in the ODRC signal. Mott insulating states are consistently observed at $V_{tg} \sim -1.37$ V in all ODRC measurements.

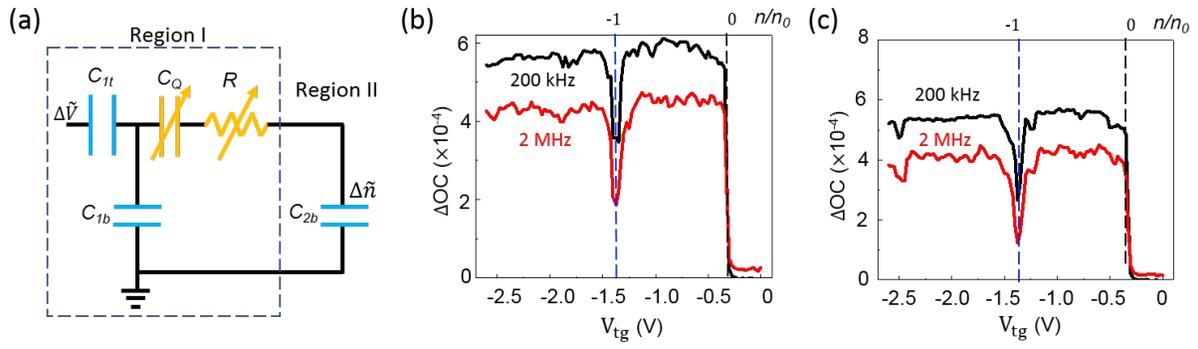

Fig. S4. (a)The diagram of the effective RC circuit for ODRC measurements. (b-c) Two additional ODRC measurements at different excitation frequencies. The hole-doped Mott insulating states appear in all measurements.

## S5. 2D finite element analysis (FEA).

We performed finite-element analysis[S5] to simulate the MIM response to 2D sheet conductance using the commercial software COMSLO 4.4. Fig. S5a shows the schematic of tip-sample geometry, where the tip is on top of the monolayer $WS_2$. The shape of the tip is defined by the following parameters: diameter $d = 100$ nm, half-cone angle $\theta = 10°$. The parameters of the device are as follows. The thicknesses of top and bottom hBN are 11 nm and 16 nm, respectively, and the dielectric constant of hBN is 4.2 [S4]. The $WS_2$ monolayer is 1 nm in thickness, with a dielectric constant of 3.5[S6]. The few-layer bottom graphene layer is ~ 4 nm thick, and its dielectric constant at GHz frequency is ~ 4 [S7]. The conductivity of the few-layer graphene is on the order of $10^4$ S/m. The substrate is Si with 90 nm $SiO_2$ whose dielectric constant is ~ 3.9 [S8].

The simulated MIM signals as a function of the sheet conductance of $WS_2$ are presented in Fig. 2b in the main text. Fig. S5b shows the quasi-static potential distribution at $\sigma_{WS_2} = 10^{-9}$ S·sq and $10^{-5}$ S·sq. Without any doping, the $WS_2$ layer is highly resistive, and there is virtually no charge screening effect from the $WS_2$ layer. In contrast, when the $WS_2$ layer is doped and becomes very conducting ($10^{-5}$ S·sq), the electric field is effectively screened by the $WS_2$ layer.

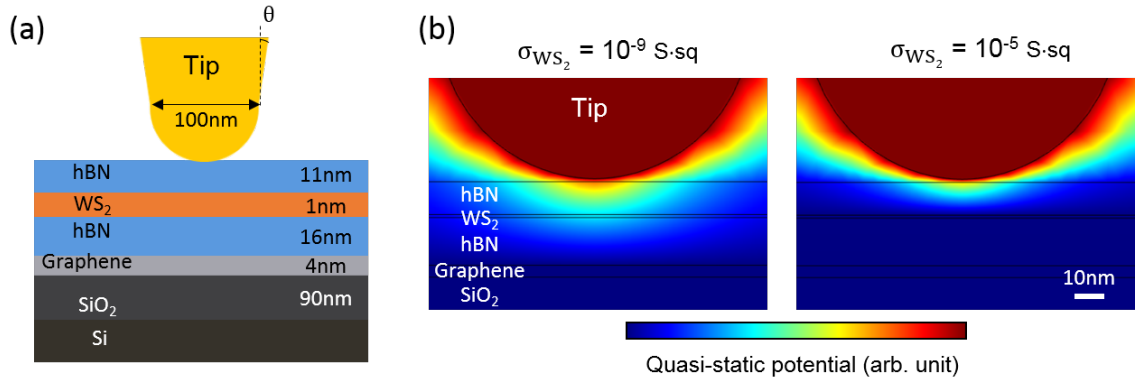

Fig. S5. (a) Schematic of tip-sample geometry for the FEA simulation. (b) Quasi-static potential distributions at $\sigma_{WS_2} = 10^{-9}$ S·sq (left) and $10^{-5}$ S·sq (right).

## S6. A complete set of the MIM data.

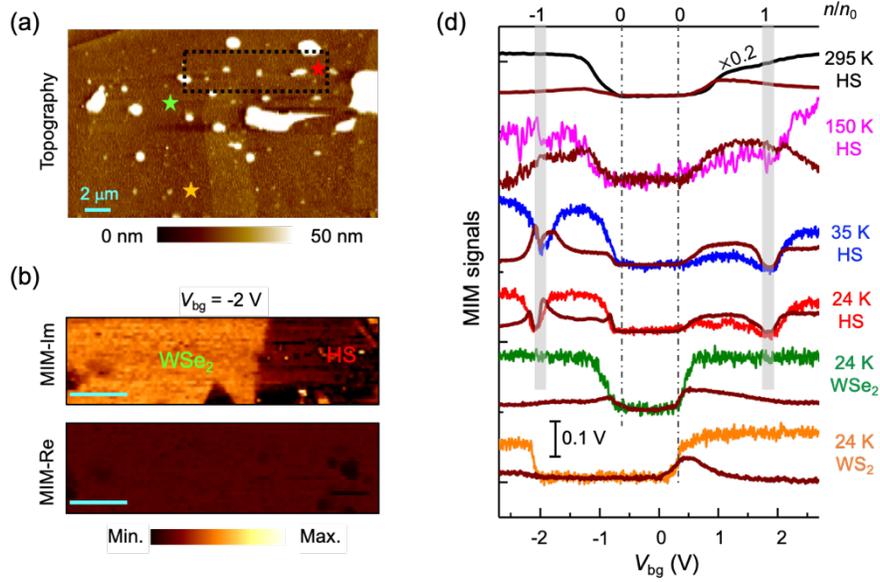

Fig. S6. (a) Topographic image of the TMD device, where the black dashed rectangle highlights the scanning area in (b,c). The red, green, and orange stars marked the locations where MIM curves (shown in (d)) were measured for heterostructure (HS), $WSe_2$ and $WS_2$ respectively. (b) MIM-Im and MIM-Re images were taken at $V_{bg}$ = -2V. (c) MIM signals as a function of $V_{bg}$ at the HS, $WSe_2$, and $WS_2$ regions marked in (a) at 24K. The MIM curves for three additional temperatures at 35K, 150K, and 295K are also shown for HS (red star). The dark red lines are the corresponding MIM-Re signals. Although there is minimal MIM-Re signal when the sample is either very conducting or insulating, the dips in MIM-Re signals indeed correspond to the correlated states. The shaded areas indicate the Mott insulating states at $n/n_0 = \pm 1$. The dash-dotted lines denote the charge neutral points, -0.7 V for the holes side and +0.3 V for the electrons side.

## S7. Correlated insulating states in a near-zero twist angle WSe$_2$/WS$_2$ device.

We investigated another WSe$_2$/WS$_2$ moiré superlattice device at 20K, where the twist angle is near zero-degree. Other parameters of the device, such as the thickness of hBN, FLG, are similar to the one presented in the main text. Fig. S7a shows the optical microscope image of the device with a relatively large field of view, where the FLG back gate and contact are marked. Fig. S7b shows the zoom-in optical picture of the TMD moiré heterostructure, which is highlighted by the white dashed rectangle in (a). The absorption spectrum at zero gate voltage at 10 K (Fig. S7c) shows three prominent peaks come from the splitting of the WSe$_2$ A exciton, which demonstrates the presence of the moiré potential. Fig. S7d shows the measured MIM-Re signals as a function of the back-gate voltage and filling factors at the point marked by the red star in (Fig. S7b). The dips in MIM-Re signals suggest the insulating states. Besides the Mott states at half filling ($n/n_0 = \pm 1$) on both electron and hole sides, we also observed insulating behaviors at full filling ($n/n_0 = \pm 2$) of the moiré band, and at -2/3 filling on the hole side. MIM-Im signals are too noisy and were not recorded during the measurements.

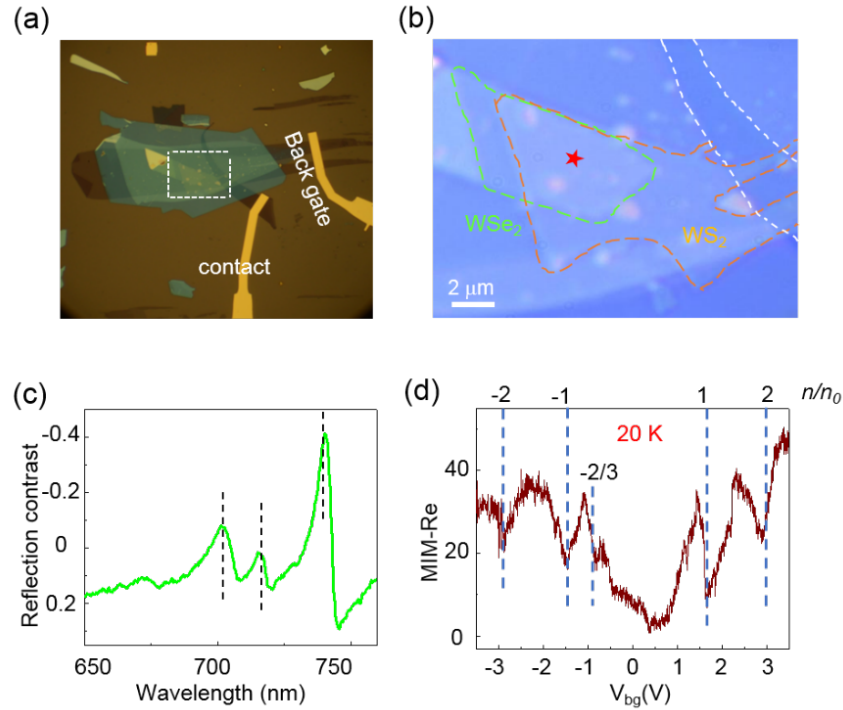

Fig. S7. (a) Optical microscope image of the device. (b) Zoom-in view of the TMD moiré heterostructure outlined by the white rectangle in (a). The FLG, WSe$_2$, and WS$_2$ regions are indicated by white, green, and orange contours. (c) Absorption spectrum at zero back-gate voltage. The dashed lines indicate the peaks in reflectance due to the splitting of the $WSe_2$ A exciton. (d) MIM-Re signals as a function of the back-gate voltage and filling factors measured at the point marked by the red star in (b). The MIM measurement was performed at 20 K.